\documentclass[aps,prd,twocolumn,english,nofootinbib]{revtex4-1}
\usepackage[T1]{fontenc}
\usepackage[latin9]{inputenc}
\usepackage{natbib}
\usepackage{color}
\usepackage{url}
\usepackage{babel}
\usepackage{array}
\usepackage{float}
\usepackage{amsmath}
\usepackage{ulem}
\usepackage{stmaryrd}
\usepackage{stackrel}
\usepackage{graphicx}
\usepackage{tikz}
\usepackage{multirow}
\usepackage{comment}
\usepackage[toc,page]{appendix}
\newcommand{\change}[1]{{#1}}
\usetikzlibrary{shapes.geometric, arrows}

\tikzstyle{startstop} = [cylinder, rounded corners, minimum width=1.5cm, minimum height=0.5cm, text width=0.8cm, text centered, draw=black, fill=blue!30, shape border rotate=90]
\tikzstyle{io} = [trapezium, trapezium left angle=70, trapezium right angle=110, minimum width=4cm, minimum height=1cm, text width=2.5cm, text centered, draw=black, fill=blue!30]
\tikzstyle{process} = [rectangle, text width=3cm, minimum width=3cm, minimum height=1cm, text centered, draw=black, fill=yellow!30]
\tikzstyle{processl} = [rectangle, text width=3cm, minimum width=3cm, minimum height=1cm, text centered, draw=black, fill=green!30]
\tikzstyle{postprocess} = [rectangle, text width=3.8cm, minimum width=4cm, minimum height=1cm, text centered, draw=black, fill=yellow!30]
\tikzstyle{postprocessl} = [rectangle, text width=3.8cm, minimum width=4cm, minimum height=1cm, text centered, draw=black, fill=green!30]
\tikzstyle{exprocess} = [rectangle, text width=3cm, minimum width=3cm, minimum height=1cm, text centered, draw=black, fill=red!30]
\tikzstyle{exprocessl} = [rectangle, text width=3.8cm, minimum width=4cm, minimum height=1cm, text centered, draw=black, fill=red!30]
\tikzstyle{inprocess} = [rectangle, text width=2.5cm, minimum width=3cm, minimum height=1cm, text centered, draw=black, fill=green!30]
\tikzstyle{decision} = [diamond, minimum width=1.5cm, minimum height=0.7cm, text width=2.08cm, text centered, draw=black, fill=green!30]
\tikzstyle{arrow} = [thick, ->, >=stealth]
\tikzstyle{arrowboth} = [thick, <->, >=stealth]

\usepackage[unicode=true,pdfusetitle,
 bookmarks=true,bookmarksnumbered=false,bookmarksopen=false,
 breaklinks=false,pdfborder={0 0 1},backref=false,colorlinks=true]{hyperref}
\hypersetup{
 linkcolor=blue,urlcolor=blue,citecolor=blue}
\makeatletter
\providecommand{\tabularnewline}{\\}
\usepackage{blindtext}

\makeatother
\begin{document}
\title{Sky localization of gravitational waves from eccentric binaries}
\author{Souradeep Pal}
\email{sp19rs015@iiserkol.ac.in}

\affiliation{Indian Institute of Science Education and Research Kolkata, Mohanpur,
Nadia - 741246, West Bengal, India}
\begin{abstract}
We demonstrate that the orbital eccentricity in compact binary mergers can be used to improve their sky localization using gravitational wave observations. Existing algorithms that conduct the localizations are not optimized for eccentric sources. We use a semi-Bayesian technique to carry out localizations of simulated sources recovered using a matched-filter search. Through these simulations, we find that if a non-negligible eccentricity is obtained during the detection, an eccentricity-optimized algorithm can significantly improve the localization areas compared to the existing methods. We also lay out the foundation for an eccentric early-warning system using the matched-filter search. The potential impact on the early-warning localization is investigated. We indicate a few possible cases of improvements while accounting for eccentricity toward any detectable eccentric neutron star binaries in the forthcoming observing scenarios of ground-based detectors. Improved localizations can be useful in effectually utilizing the capabilities of the follow-up facilities.
\end{abstract}
\date{\today}
\maketitle
\section{Introduction}\label{sec:intro}
\begin{figure*}[t]
\begin{centering}
\includegraphics[scale=0.65]{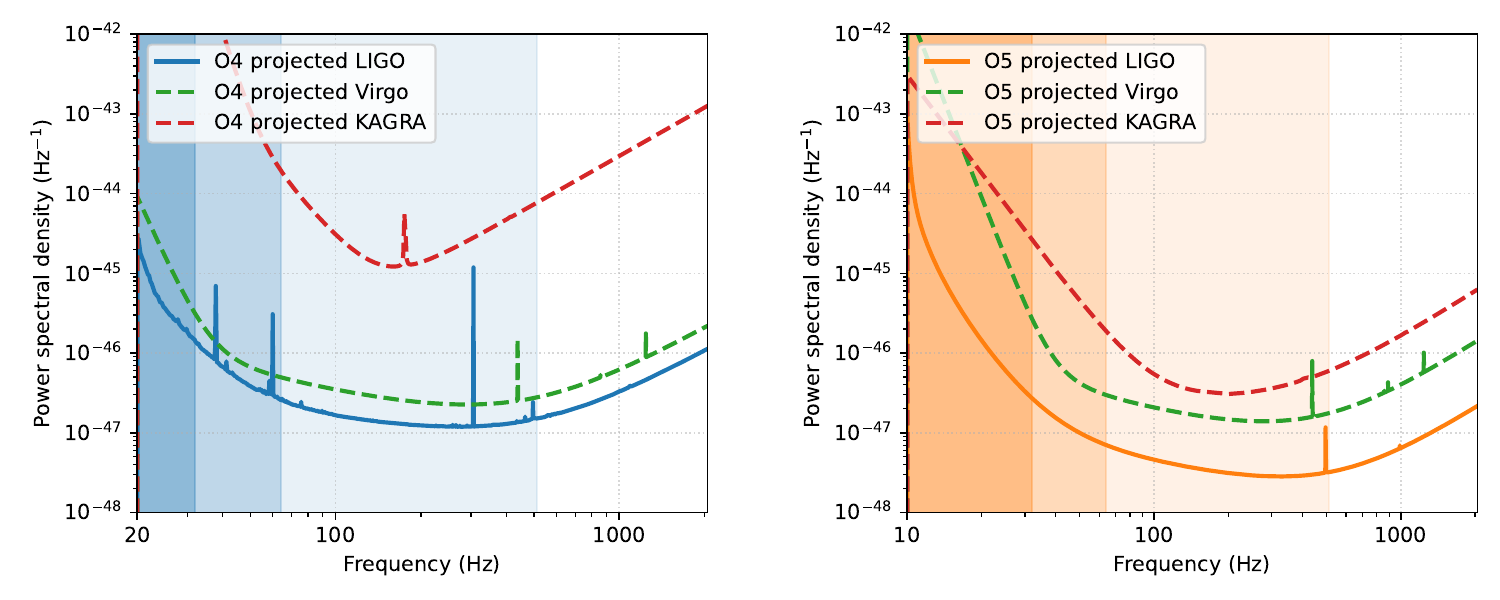}
\par\end{centering}
\caption{The projected power spectral densities (PSDs) of the Advanced LIGO in O4 and O5 are used for the simulations. Advanced Virgo and KAGRA are expected to be instrumental in detecting or refining the localizations of sources while advancing their sensitivities. Here all investigations are carried out with simulated Gaussian noise. We use a low-frequency cutoff of 20 Hz and 10 Hz respectively for the simulated O4 and O5 scenarios. The shaded regions mark the frequency ranges used for the simulated searches in this work.}\label{fig:psd}
\end{figure*}
Till now gravitational wave signals from two binary neutron star (BNS) mergers have been reported~\citep{abbott2017gw170817,Abbott_2020}. Accurate and rapid localization of such sources in the sky increases the chances of finding any multi-messenger counterparts~\citep{singer2016rapid,Chen_2017,Fairhurst_2018, Hu_2021,Chatterjee_2023}. The accuracy of the localizations benefits from adding new detectors to the global network of gravitational wave detectors and/or by upgrading the existing ones~\citep{Fairhurst_2014,Saleem_2021,pandey2024criticalroleligoindiaera}. It can also be supported by optimizing the localization algorithms toward the expected population of sources~\citep{Kapadia_2020,10.1093/mnras/stac715}.

Although a non-neglibible orbital eccentricity associated with a binary neutron star inspiral is expected to be a rare case, there are considerable efforts toward detecting any such systems~\citep{PhysRevD.87.127501,Nitz_2020,pal2023swarm,dhurkunde2023searcheccentricnsbhbns}. There are also several attempts to measure the eccentricities of LIGO-Virgo-KAGRA events, for example~\citep{10.1093/mnras/staa2120,Romero-Shaw_2020,gayathri2022eccentricity,gupte2024evidence,wu2020measuring,romero2019searching,Romero-Shaw_2021,Romero-Shaw_2022,10.1093/mnrasl/slaa084,PhysRevD.107.064024,gamba2023gw190521,iglesias2022reassessing}. As the low-frequency sensitivities of the detectors improve, the chances of observing any eccentric binary inspiral become prominent and promising, especially with the next-generation gravitational wave detectors~\citep{dhurkunde2023searcheccentricnsbhbns,Nitz_2021}. Studies indicating various observational prospects with eccentric binaries can be found in the literature, for example in~\citep{PhysRevD.109.104041,PhysRevLett.129.191102,PhysRevD.96.084046,yang2024advantageearlydetectionlocalization}.

The data taking campaigns of the Advanced LIGO, Advanced Virgo and KAGRA are split into months-long observation periods~\citep{abbott2020prospects,aasi2015advanced,capote2024advanced,buikema2020sensitivity,martynov2016sensitivity,abbott2016gw150914,acernese2014advanced,PhysRevLett.123.231108,akutsu2021overview}. When the detectors are observing, several automated analyses search through the recorded data for potential astrophysical events in real-time~\citep{messick2017analysis,Aubin_2021,nitz2018rapid,PhysRevD.86.024012,PhysRevD.72.122002,PhysRevD.95.104046,PhysRevD.110.104034}. Whenever a candidate event is detected with a significance above a set threshold, another automated analysis is triggered that inspects the detection candidate for consistency across multiple analyses and launches the sky localization. Information relevant for immediate electromagnetic (EM) follow up of the event are combined and an automated message is prepared. A preliminary notice is sent out within seconds or minutes to engage astronomers from the broader community~\citep{Chaudhary_2024}.

For a sufficiently long and loud signal, there are substantial chances of observing the signal even before it reaches the merger~\citep{Cannon_2012}. To enable such observations, dedicated analyses process the data as soon as they are available and intermittently checks for any potential candidate without waiting for the entire signal to pass through the detectors~\citep{Nitz_2020_1,Sachdev_2020,Kovalam_2022}. Such analyses, limited in time and frequency, can lead to early-warnings of the merger in the signals. If a true early-warning candidate is found, it must be succeeded soon by a consistent full-time, full-frequency-bandwidth candidate and similar follow-up processes are carried out as described earlier. The prospects for early-warning observations improve with better low-frequency sensitivities of the instruments when more early-inspiral cycles from a binary lie within the sensitive frequency band.

Since orbital eccentricity has a prominent effect primarily in the low signal frequencies, here we investigate its effects on the localization during the detection of a binary inspiral. Through simulations, we set up a scenario where mock signals from eccentric BNSs are recovered from simulated data using an eccentric matched-filter search and are then localized using an eccentricity-optimized localization algorithm. We primarily compare the outcome of the eccentric analysis with an, otherwise same, non-eccentric analysis. We show that in the presence of a non-negliglible eccentricity in the signals, the localizations can improve significantly. The improvement is expected to be more pronounced for lighter binaries where the number of low-frequency inspiral cycles in the sensitive frequency band is larger. In the absence of a residual eccentricity in the signals, the accuracy of the method falls back to that of the non-eccentric localizations on average.

The basic principle of the localization methods used in this work is to assume that the \textit{intrinsic} source parameters obtained from matched-filter searches are the true values themselves~\footnote{Alternately, this assumption can be lifted toward a full Bayesian parameter estimation, thus increasing the parameter dimensionality and the cost of computing the likelihoods. Several optimized approaches already implement this idea without including eccentricity~\citep{Morisaki_2020,10.1093/mnras/stab2977,PhysRevLett.127.241103,Pathak_2024}.}. In practice, the detection algorithms provide point estimates of these parameters that can carry large uncertainties, e.g. for eccentric sources~\citep{pal2023swarm}. Ideally, the set of true source parameters maximizes the recovered signal-to-noise ratio (SNR), however, as long as the any other combination forming a detection candidate, which nearly maximizes the SNR, can still be used. It has been shown that such a combination can be utilized for the purpose of rapid localization, laying the basis for the \texttt{BAYESTAR} algorithm, which produces skymaps that are almost as accurate as that of a full parameter estimation. This has been demonstrated for sources with small aligned-spins~\citep{singer2016rapid}.

Here we use a semi-Bayesian approach to obtain the eccentricity-optimized localizations utilizing the point estimate for the orbital eccentricity from the detection template. Thus, in essence, the localization principle here is similar to that of \texttt{BAYESTAR}, however the difference lies in the implementation. The main contrasting feature is that the likelihood sampling in our approach is performed on the flat-sky coordinates, namely the right ascension and the declination, using a stochastic sampler. The obtained samples are later reweighted and projected onto the sky. On the other hand, \texttt{BAYESTAR} deterministically samples the likelihood directly on the curved-sky using an adaptive \texttt{HEALPix} algorithm~\footnote{\url{https://healpix.sourceforge.io}}. Other differences include the choice of marginalization of the auxiliary parameters such as the luminosity distance and the inclination angle. While the current \texttt{BAYESTAR} algorithm marginalizes them to obtain the posterior probabilities on the sky, our approach stochastically samples them alongside the sky coordinates. Essentially, note that \texttt{BAYESTAR} localizations are not optimized for eccentric binaries.

The goal of this work is to explore the localization differences of the above methods toward eccentric sources with the sensitivities of the current detectors. We primarily use the projected sensitivities of the Advanced LIGO for the fourth and the fifth oberving runs (O4 and O5)~\citep{pubpsd}. These are shown in Fig.~\ref{fig:psd}. We use simulated Gaussian data to inject the simulated signals. When we compare the localizations, we assume that a detection has already been made, fulfiling a set of criteria as described later in the work. This work is organized as follows. In Section~\ref{sec:simsea}, we briefly outline the matched-filter search based on Particle Swarm Optimization (PSO)~\citep{pal2023swarm}. To efficiently tackle the intrinsic parameter space, we introduce a computationally optimized variant of the standard PSO algorithm. We first describe the algorithm in recovering simulated signals from an astrophysically agnostic population of eccentric BNSs. The signals carry small eccentricities in the presence of large aligned-spins. In Section~\ref{sec:skyloc}, we use the detection outputs of the simulated searches obtained in Section~\ref{sec:simsea} to perform the localizations. The processes used in the localizations and the expected improvements are discussed in detail. In Section~\ref{sec:earwar}, we focus our attention specifically to the early-inspiral phase of the signals. We conclude with the implications for early-warning scenarios for O4 and O5.
\begin{table*}[t]
\begin{centering}
\begin{tabular}{>{\centering}p{6cm}>{\centering}p{2.4cm}>{\centering}p{2.4cm}>{\centering}p{2.4cm}>{\centering}p{2.4cm}}
\hline 
{\footnotesize{}Parameter} & {\footnotesize{}Type} & {\footnotesize{}Distribution} & {\footnotesize{}Range} & {\footnotesize{}Recovery}\tabularnewline
\hline 
\hline 
{\footnotesize{}Primary mass, $m_{1}(M_{\odot})$} & {\footnotesize{}Intrinsic} & {\footnotesize{}Uniform} & {\footnotesize{}{[}1.0, 3.0{]}} & {\footnotesize{}{Frequentist}}\tabularnewline
{\footnotesize{}Secondary mass, $m_{2}(M_{\odot})$} & {\footnotesize{}Intrinsic} & {\footnotesize{}Uniform} & {\footnotesize{}{[}1.0, 3.0{]}} & {\footnotesize{}{Frequentist}}\tabularnewline
{\footnotesize{}Primary aligned-spin, $\chi_{1z}$(dimensionless)} & {\footnotesize{}Intrinsic} & {\footnotesize{}Uniform} & {\footnotesize{}{[}-0.95, 0.95{]}} & {\footnotesize{}{Frequentist}}\tabularnewline
{\footnotesize{}Secondary aligned-spin, $\chi_{2z}$(dimensionless)} & {\footnotesize{}Intrinsic} & {\footnotesize{}Uniform} & {\footnotesize{}{[}-0.95, 0.95{]}} & {\footnotesize{}{Frequentist}}\tabularnewline
{\footnotesize{}Orbital eccentricity, $\varepsilon$ (dimensionless)} & {\footnotesize{}Intrinsic} & {\footnotesize{}Uniform} & {\footnotesize{}{[}0, 0.15{]}} & {\footnotesize{}{Frequentist}}\tabularnewline
{\footnotesize{}Right ascension, $\alpha$ (rad)} & {\footnotesize{}Extrinsic} & {\footnotesize{}Uniform} & {\footnotesize{}{[}0, $2\pi${]}} & {\footnotesize{}{Bayesian}}\tabularnewline
{\footnotesize{}Declination, $\delta$ (rad)} & {\footnotesize{}Extrinsic} & {\footnotesize{}Uniform} & {\footnotesize{}{[}$-\pi/2$, $\pi/2${]}} & {\footnotesize{}{Bayesian}}\tabularnewline
{\footnotesize{}Luminosity distance, $D_{L}$ (Mpc)} & {\footnotesize{}Extrinsic} & {\footnotesize{}Uniform} & {\footnotesize{}{[}50, 160{]}} & {\footnotesize{}{Bayesian}}\tabularnewline
{\footnotesize{}Inclination angle, $\iota$ (rad)} & {\footnotesize{}Extrinsic} & {\footnotesize{}Uniform} & {\footnotesize{}{[}0, $\pi${]}} & {\footnotesize{}{Bayesian}}\tabularnewline
{\footnotesize{}Polarization, $\psi$ (rad)} & {\footnotesize{}Extrinsic} & {\footnotesize{}Uniform} & {\footnotesize{}{[}0, $2\pi${]}} & {\footnotesize{}{Marginalized}}\tabularnewline
{\footnotesize{}Coalescence phase, $\varphi$ (rad)} & {\footnotesize{}Extrinsic} & {\footnotesize{}Uniform} & {\footnotesize{}{[}0, $2\pi${]}} & {\footnotesize{}{Marginalized}}\tabularnewline
{\footnotesize{}Coalescence time, $t_{c}$ (seconds)} & {\footnotesize{}Extrinsic} & {\footnotesize{}Uniform} & {\footnotesize{}{[}$t_{c}-0.1$, $t_{c}+0.1${]}} & {\footnotesize{}{Marginalized}}\tabularnewline

\hline 
\end{tabular}
\par\end{centering}
\caption{Summary of the signal parameters used to generate and recover injections in this work. Here the recovery space is the same as that used for carrying out the injections. An arbitrary GPS time offset is added to $t_{c}$ for an injection. A sufficient duration of noise segment is used to accomodate the injected signals with a given low-frequency cutoff. The eccentricity is defined at the low-frequency cutoff. Note that the \texttt{TaylorF2Ecc} waveform model does not implement the mean anomaly parameter.\label{tab:injpar}}
\end{table*}
\section{Simulated searches}\label{sec:simsea}
This section briefly describes the processes used to recover simulated signals from simulated data. We create a simulated population of BNS systems with the signal parameters given in Table~\ref{tab:injpar}. To generate the signals, we use the \texttt{TaylorF2Ecc} waveform model~\citep{moore2016gravitational}. For any given detector, a signal is projected from the random sky location and is deposited in Gaussian noise colored with the simulated PSD for the detector. To recover injections in this work, we broadly make the following two changes to the PSO-based detection framework described in~\citep{pal2023swarm}. These changes are made primarily to simplify simulating the search process and to efficiently tackle a larger search parameter space.

Given a stretch of coincident data across detectors, the PSO algorithm iteratively optimizes a quantity called the \textit{fitness function} over a desired range of the search parameters. Here the search parameters are the intrinsic parameters of the sources. At any given iteration, the algorithm proposes a set of template points, calculates the fitness function at these points and evaluates a new set of template points for the next iteration. Here, for each template point proposed, the template waveform is generated, match-filtered with the strain data and the maximum SNR per detector is identified. The algorithm iteratively optimizes the quadrature sum of the detector SNRs over the search space for each injection. Because the injections are in Gaussian noise, we proceed with a single \textit{trigger} per detector for the data stretch which has the largest SNR. Thus for simplicity, we skip the rigorous process of generating and ranking of multiple possible event candidates for the injections. Note that in this work we are primarily interested in the localization of detected events instead of the detection process itself.
\begin{figure}[b]
\begin{centering}
\includegraphics[scale=0.60]{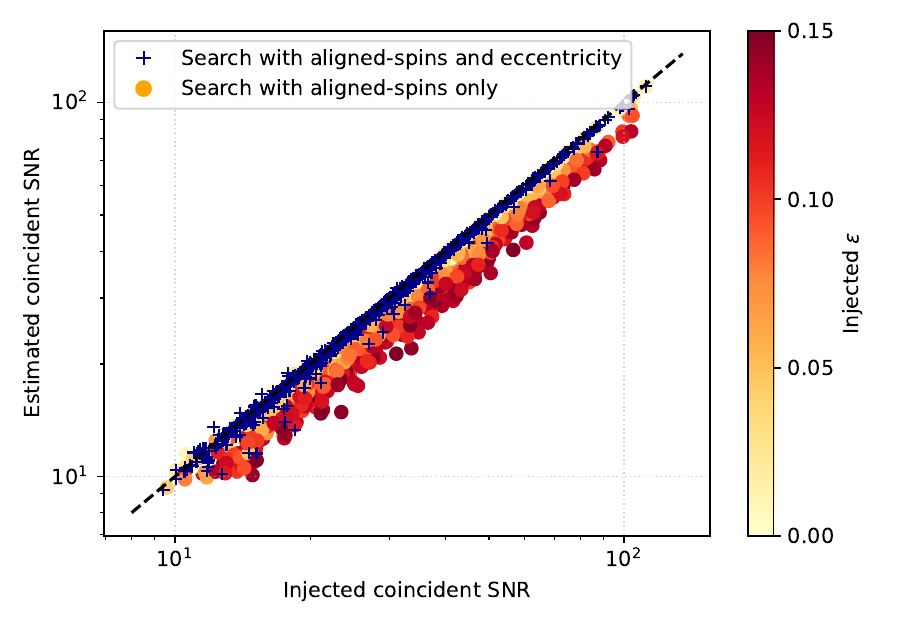}
\par\end{centering}
\caption{The coincident SNRs recovered by the PSO searches for simulated signals in the HL network having O4-like sensitivities. The searches are carried out in the frequency range of 20-512 Hz using the same datasets containing signals from spin-aligned eccentric BNSs immersed in Gaussian noise. The injected SNRs are computed at the injected parameters with a high-frequency cutoff of 2048 Hz. We use a fixed template-sampling for both the searches as described in the text. The eccentric search recovers all the injections that are also recovered by the aligned-spin search. The commonly found injections are shown here.}\label{fig:fbsnrs}
\end{figure}

To tackle the search parameter space efficiently, we implement a nature-inspired \textit{survival of the fittest} strategy on the PSO algorithm as described below. We observe that to cover larger portions of parameter space, the algorithm requires overall greater template-sampling. Further, as the iterations progress, we notice that the \textit{particles} with the smallest fitness values barely contribute to the swarm-intelligence. Thus, it is wasteful to evaluate the expensive fitness function for the trailing particles and are terminated at regular intervals. So the general idea is to begin with a swarm having a sufficiently large number of particles and reduce the swarm size by approximately half at each one-thirds of the total iterations. Here, starting with a fixed number of 7200 particles and 18 iterations, a total of 75600 template points are calculated~\footnote{$7200\times6+3600\times6+1800\times6$; on the contrary, the number of particles is fixed at every iteration in the standard PSO approach. For a given population of simulated sources, we empirically preset the template-sampling to tolerate a nominal value for the average loss of the recovered coincident SNRs from that of the injected SNRs. In this study, we observe this value to be approximately 1.75\% in the full-frequency bandwidth search corresponding to the blue pluses in Fig.~\ref{fig:fbsnrs}.}. We find that this strategy is more effective than the standard algorithm in optimizing the dynamic template placement while tackling a reasonably large parameter space.

We adapt the above changes to recover injections using the following two searches- aligned-spins with eccentricity and aligned-spins only. The two searches use identical configuration except for the search dimension of eccentricity. Note that the template-sampling is also preset at a fixed value for both the searches. A similar framework of analysis is demonstrated earlier in~\citep{pal2023swarm}. The estimated coincident SNRs plotted against the injected coincident SNRs for the recovered injections are shown in Fig.~\ref{fig:fbsnrs}. We note the difference in the SNRs recovered by the two searches for any given injection with a general dependence on its injected value of eccentricity.

After the iterations are complete, the algorithm delivers an optimized template point from the search paramater space for each injection. This is regarded as the best-matched template consistent for the given detector network (HL), nearly maximizing the detector SNRs. Injections that have individual detector SNRs of at least 5 and form a coincidence within a time-window of 15 ms are regarded as the recovered \textit{event} candidates~\citep{pal2023swarm}. Such injections are taken up by the sky-localization algorithm as described in the next section.
\section{Localization algorithm}\label{sec:skyloc}
In the previous section, we described the recovery of simulated signals as coincident events. The localization of a recovered event is performed as described below. To compare the eccentric and non-eccentric localizations, we pre-select the injections that are recovered exactly as two-detector coincidences by both the eccentric and aligned-spin searches. Thus we skip the cases of the sub-threshold signals in a detector resulting from a given search.
\begin{figure}[t]
\begin{centering}
\includegraphics[scale=0.65]{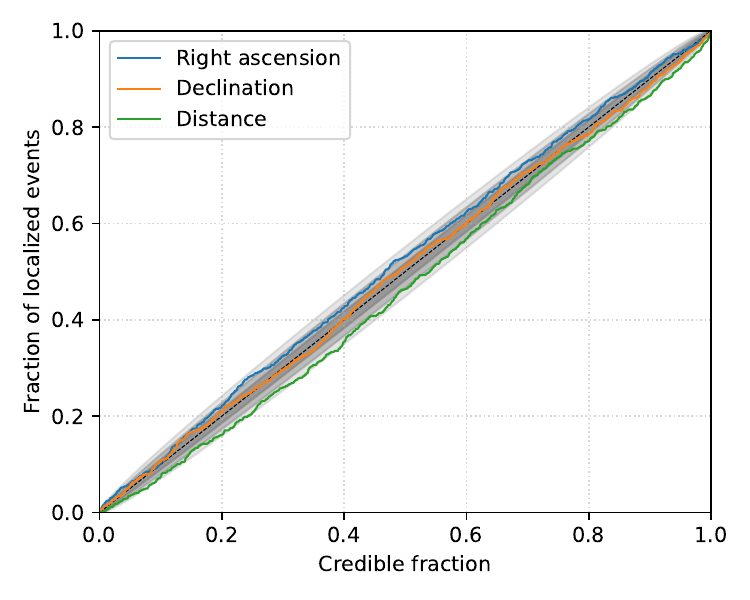}
\par\end{centering}
\caption{PP-plots for the 3D localization parameters shown for the eccentric analysis of the simulated sources described in the Table.~\ref{tab:injpar}. These are obtained without any non-trivial correction factor (or with a corresponding default value of 1), which is described later in the text. The gray regions represent the $1$, $2$, and $3\sigma$ uncertainty estimates obtained from beta distribution.}\label{fig:ppsam}
\end{figure}
\begin{figure*}[t]
\begin{centering}
\includegraphics[scale=0.65]{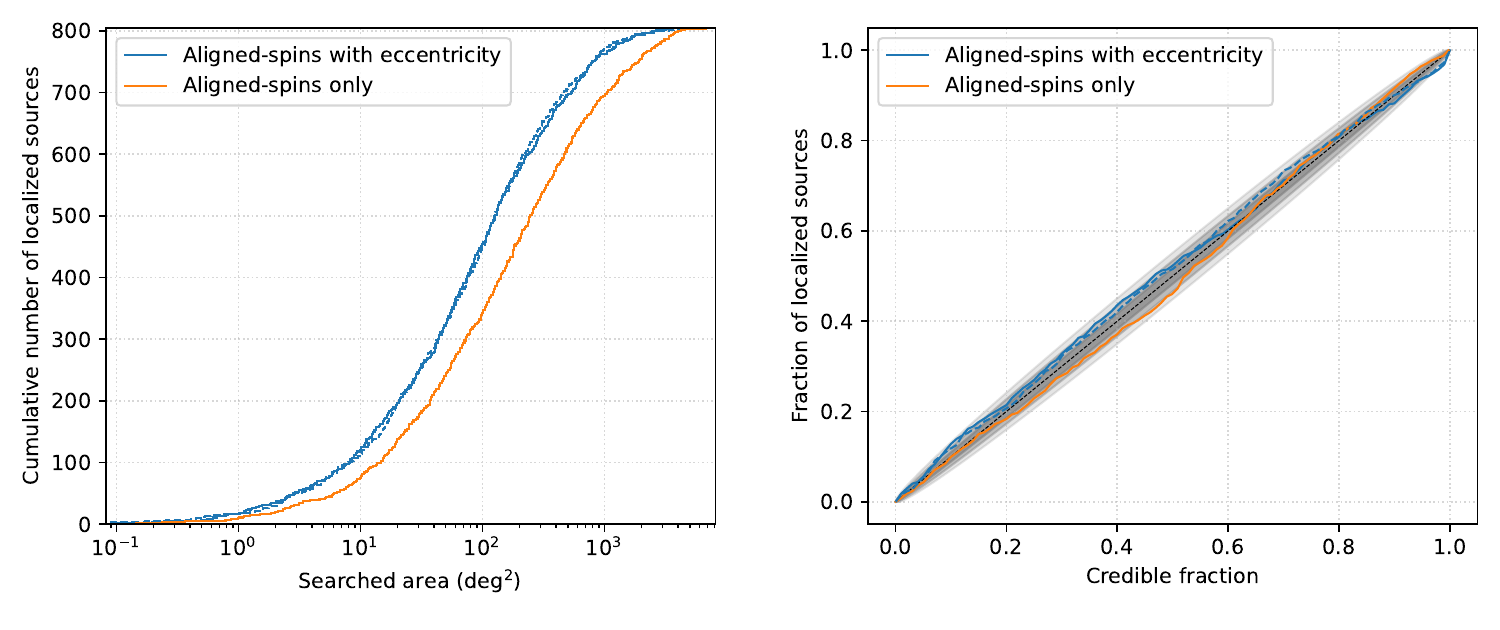}
\par\end{centering}
\caption{Accuracy (left) and self-consistency (right) of the localization areas compared for the simulated eccentric sources. The solid curves are obtained using the estimated intrinsic source parameters with the PSO-based searches. The dashed curves are obtained using the injected source parameters. The probability plots on the right show that the credible sky areas are consistent at a 99.7\% confidence.}\label{fig:prodsky}
\end{figure*}
\begin{figure}[b]
\begin{centering}
\includegraphics[scale=0.65]{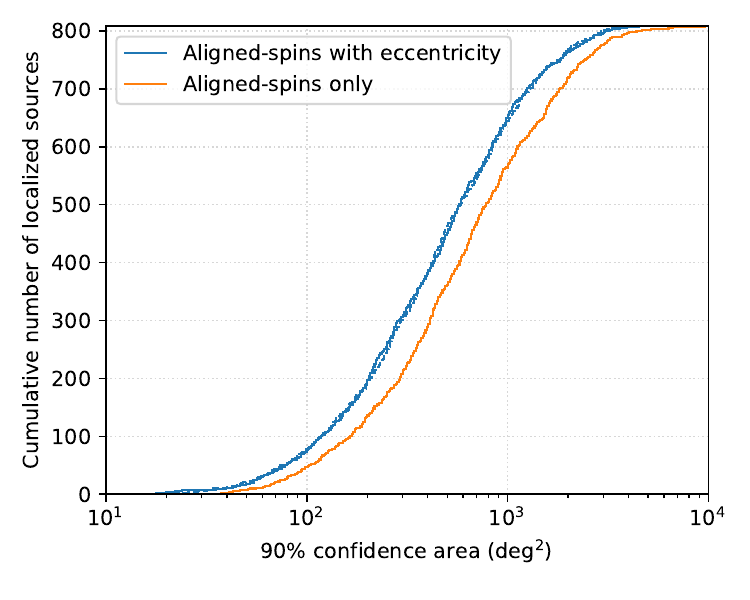}
\par\end{centering}
\caption{Cumulative historgrams showing the distribution of a credible (90\%) area for the eccentric injections obtained with the eccentric and non-eccentric analyses.}\label{fig:areaHL}
\end{figure}
\begin{figure*}[t]
\begin{centering}
\includegraphics[scale=0.11]{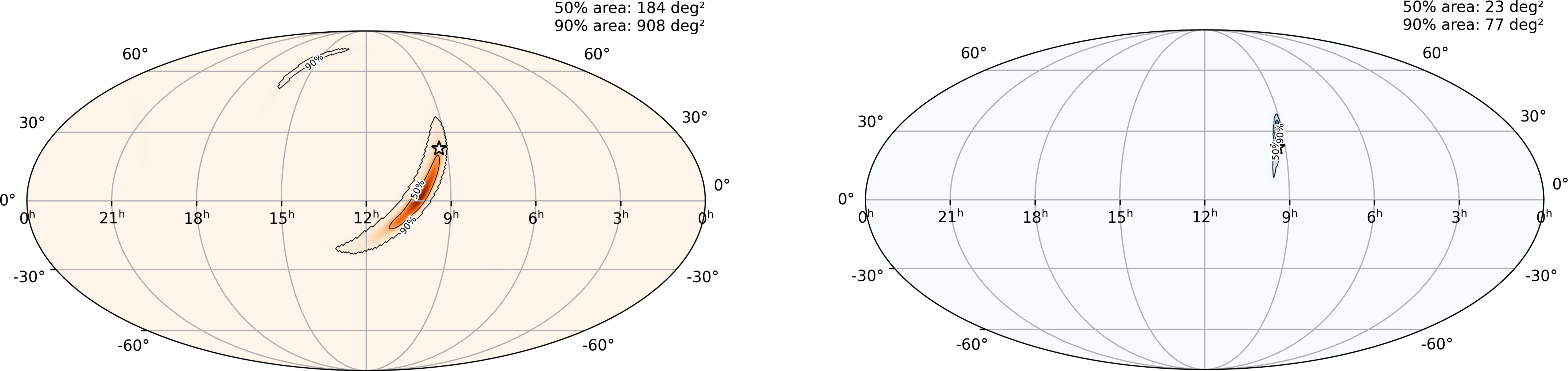}
\par\end{centering}
\caption{Skymaps for an eccentric injection obtained with the aligned-spins only (left) and the eccentricity-optimized (right) localization algorithms. The injection is a randomly chosen (1.93, 1.88) $M_{\odot}$ source located at about 65 Mpc, having an eccentricity of 0.1 and nominal aligned-spin values. The localizations are computed following sections Sec.~\ref{sec:simsea} and~\ref{sec:skyloc} with simulated HLV network. The true location of the simulated source is marked with a star. Notably, we observe that the localization areas at the given credible levels are significantly improved with the eccentricity-optimized algorithm. Further, note the existence of two probability modes (left) compared to a single mode (right) at 90\% credible level. Tracking skymaps with multiple separated modes could be resource-demanding process in the EM follow-up communities.}\label{fig:ful_sky}
\end{figure*}
\begin{figure}[b]
\begin{centering}
\includegraphics[scale=0.65]{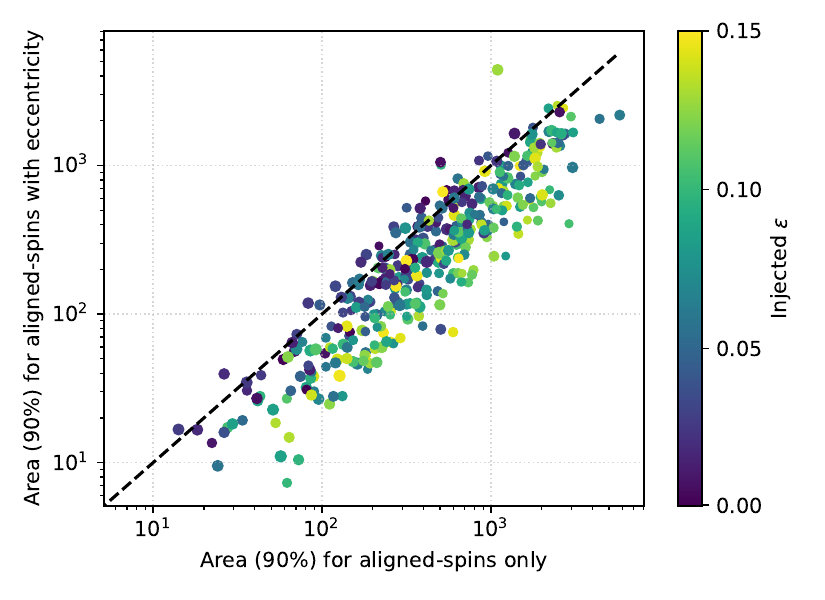}
\par\end{centering}
\caption{Dependence of the localization improvement on the eccentricity in terms of the 90\% confidence area for the eccentric injections obtained with the eccentric and non-eccentric analyses described in the text.}\label{fig:skyecc}
\end{figure}
Once a detection has been made, the strain data around the event time from the observing detectors become of particular interest. Here, the localization algorithm receives a sufficient duration of the strain data containing the entire in-band signal determined by its chirp time. The chirp time is calculated using the chirp mass estimate of the source during its detection and the low-frequency cutoff. Note that the PSO search provides a detection candidate corresponding to the template point that optimizes the SNR computed for the detector network. Given the above information, it is possible to assign a probability that the source is located at any given part of the sky. As indicated in~\citep{singer2016rapid}, the localization uncertainty is only semi-dependent on the uncertainties of the intrinsic source parameters. Thus we conduct a semi-Bayesian analysis by fixing the intrinsic parameters to that obtained as point estimates from the optimized template~\footnote{Both the detection and the localization methods reported in this work will need to account for the earth's rotation effects during a longer inspiral, depending upon the low-frequency cutoff of the analyses. However, note that the output from only a single (optimized) template bridges the two algorithms. Thus, as long as the detection algorithm reports an optimal candidate, a rotation-optimized localization algorithm within the current scheme is also expected to be optimal. We skip such investigations for a future study.}.

After fixing all the intrinsic parameters, we are left with the extrinsic parameters as shown in the Table.~\ref{tab:injpar}. We stochastically draw likelihood samples from the following set of parameters,
\begin{equation}
\boldsymbol{\theta}=\{\alpha, \delta, D_{L}, \iota\}.\label{eq:params}
\end{equation}
The Bayes' rule allows us to compute the posterior probabilities of a parameter ($\theta$), given the likelihood of observing the data ($d$) for the given $\theta$ and a prior knowledge about $\theta$,
\begin{equation}
p(\theta\,|\,d)\sim{p}(d\,|\,\theta)\:p(\theta).
\end{equation}
Note that the parameters given in Eq.~\ref{eq:params} are intended to aid electromagnetic follow-up efforts. Since the injected population has uniformly distributed signal parameters in their respective domains, here we use a uniform prior for each of the parameters. When the injected population is more astrophysically informed, it may be reasonable to use a similar distribution for the prior, see for example~\citep{Fairhurst_2018}. This can be achieved with minimal changes in the analysis configuration. Here we choose to marginalize the remaining extrinsic parameters. When the form of the signal can be adequately represented in the dominant (2, 2) mode, the uniform prior over $\varphi$ leads to a simple analytic marginalization of the posterior. The marginalizations over $t_{c}$ and $\psi$ use a uniform grid of $1000$ points in their respective prior spaces and require interpolation of the likelihood followed by explicit numeric marginalization over the resulting array of points. The posterior samples are obtained with the \texttt{dynesty} sampler~\citep{10.1093/mnras/staa278}, configured with \texttt{nlive} $=1000$ and \texttt{dlogZ} $=0.1$. This is accomplished within $25$ seconds using $\sim32$ CPU cores. A \textit{probability-probability} (pp)-plot derived from the posterior samples for each of the three-dimensional (3D) localization parameters is shown in Fig.~\ref{fig:ppsam}. Note that the posterior samples are obtained in the flat $\alpha-\delta$ coordinates. To compare with the localizations from \texttt{BAYESTAR}~\citep{singer2016rapid}, the samples are rendered into multi-ordered \texttt{HEALPix} format~\citep{fernique2015moc,2005ApJ...622..759G,Zonca2019}. Assigning probabilities to the \textit{pixels} tiling the sky using the distribution of samples requires an additional postprocessing step~\citep{ligo.skymap}.

To assess the accuracy of the localizations, we use a quantity called \textit{searched area} which measures the area-distance of the true source location from the most probable location on the skymap. It is calculated by adding the areas of the pixels lined up in descending order of the probabilities, starting at the most probable pixel until the pixel containing the true location is reached. 
We compare the searched area for the spin-aligned eccentric injections described in Section~\ref{sec:simsea} with the following two cases- (1) the localizations obtained with the above algorithm for the HL coincidences from the spin-aligned eccentric PSO search, and (2) the localizations obtained with \texttt{BAYESTAR} for the same HL coincidences from the aligned-spin PSO search. This is shown in Fig.~\ref{fig:prodsky} (left). We repeat case (1) by replacing the PSO estimated intrinsic source parameters with the true injected parameters. We observe that the localizations with the injected and the estimated parameters for the eccentric case fairly agree with each other. However, the localizations for the non-eccentric aligned-spin case are generally less accurate than the eccentric case. We attribute the improvement in accuracy to the difference in the estimated SNRs as shown in Fig.~\ref{fig:fbsnrs}. We also note that a small fraction of the injections have the localizations of $\sim$ 1~$\mathrm{deg}^{2}$ depending on factors such as the signal strength and relative orientation of the binary with respect to the detector network. We do not apparently observe large differences for them from the two analyses. Further studies are needed to investigate if these analyses are capable of resolving such finer localizations. On the other hand, we have capped our comparisons to localizations areas of $\sim$ 10000~$\mathrm{deg}^{2}$. Current typical localizations generally lie within this range, the regime where we observe greater improvements.

To assess the self-consistency of the localizations, we use the pp-plot to describe the credibility of the localization areas on the skymaps. A given credible localization patch on the skymap is determined by the group of pixels whose probabilities add up to the desired credible level, starting at the most probable pixel. The pp-plot is calculated by obtaining the fractions of the localized population containing the true location at all credible levels. This is shown in Fig.~\ref{fig:prodsky} (right). We note that each configuration of localization described above requires a correction factor on the likelihood computations to obtain a self-consistent pp-plot discussed further in Appendix~\ref{sec:locunc}. Thus, on average, the localization areas obtained with these methods can be considered reliable for the simulated population. We further note that the precision of the localization area at 90\% confidence are improved for the eccentric analysis as compared to the non-eccentric one, obtained for the commonly found injections, as shown in Fig.~\ref{fig:areaHL}. An example of the localization differences is illustrated in Fig.~\ref{fig:ful_sky}. The general dependence of the improved precision on the eccentricity for the simulated population can be seen in Fig.~\ref{fig:skyecc}. While the improvement generally scales with eccentricity, it is further dependent on several other factors. The in-band signal, which depends on the frequency cutoffs and the chirp mass, primarily dictates how critical a given value of eccentricity is. The improvement could also be dependent on the unoptimized area itself, which depends on the distance and the orientation of the source as discussed earlier. \change{A discussion on the distance estimates under different priors is presented in Appendix~\ref{sec:dist}.}
\begin{figure*}[t]
\begin{centering}
\includegraphics[scale=0.7]{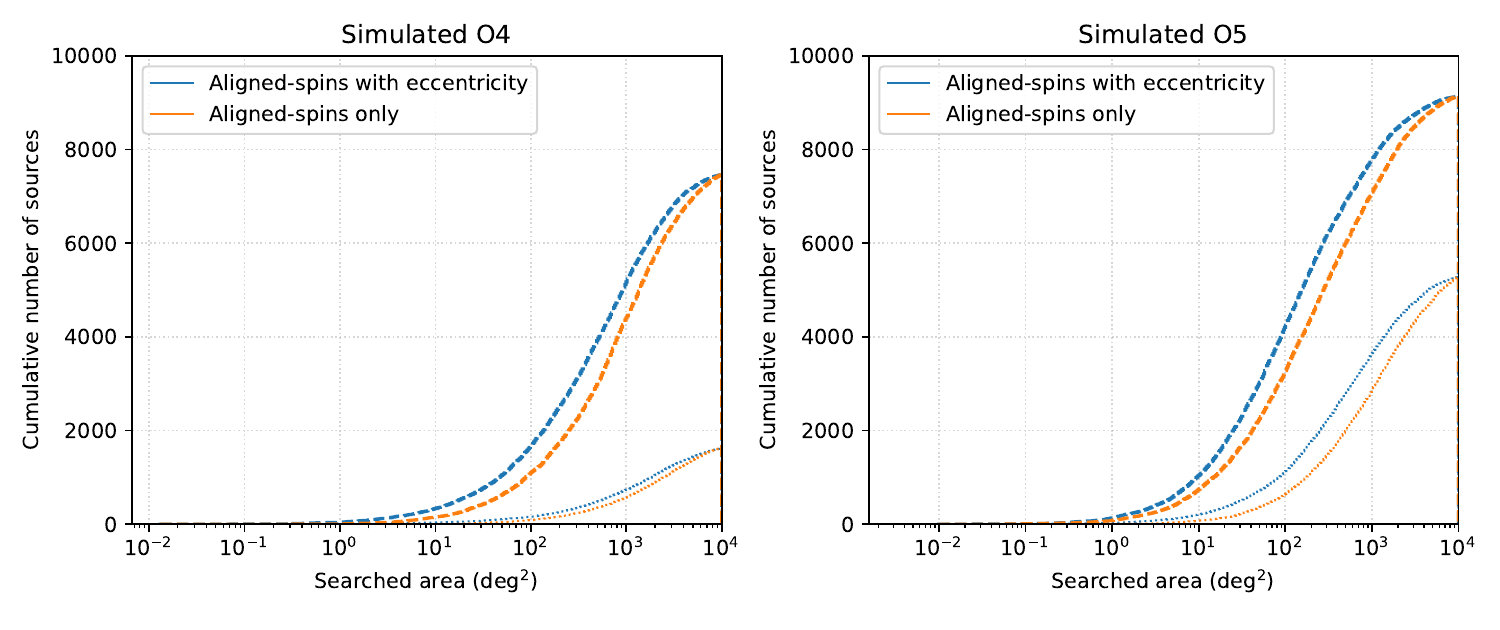}
\par\end{centering}
\caption{Localization accuracies with upper cutoff frequencies of 32 Hz (dotted) and 64 Hz (dashed) for spin-aligned eccentric injections. We observe generally prominent improvements in the eccentricity-informed localization areas while greater injections are recovered in O5 than in O4. Injections that are localized within a searched area of 10000 $\mathrm{deg}^{2}$ are shown.}\label{fig:ear_loc}
\end{figure*}
\begin{figure}[b]
\begin{centering}
\includegraphics[scale=0.7]{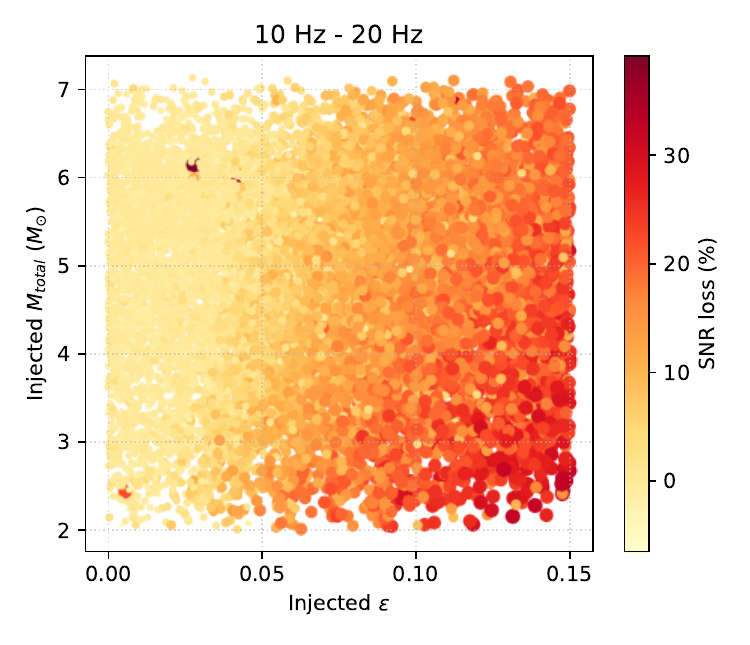}
\par\end{centering}
\caption{Variation of the loss in the recovered SNR by a simulated aligned-spin search with the injected eccentricity and the total mass of the system in O5-like sensitivity.}\label{fig:mchvsecc}
\end{figure}
\begin{figure*}[t]
\begin{centering}
\includegraphics[scale=0.6]{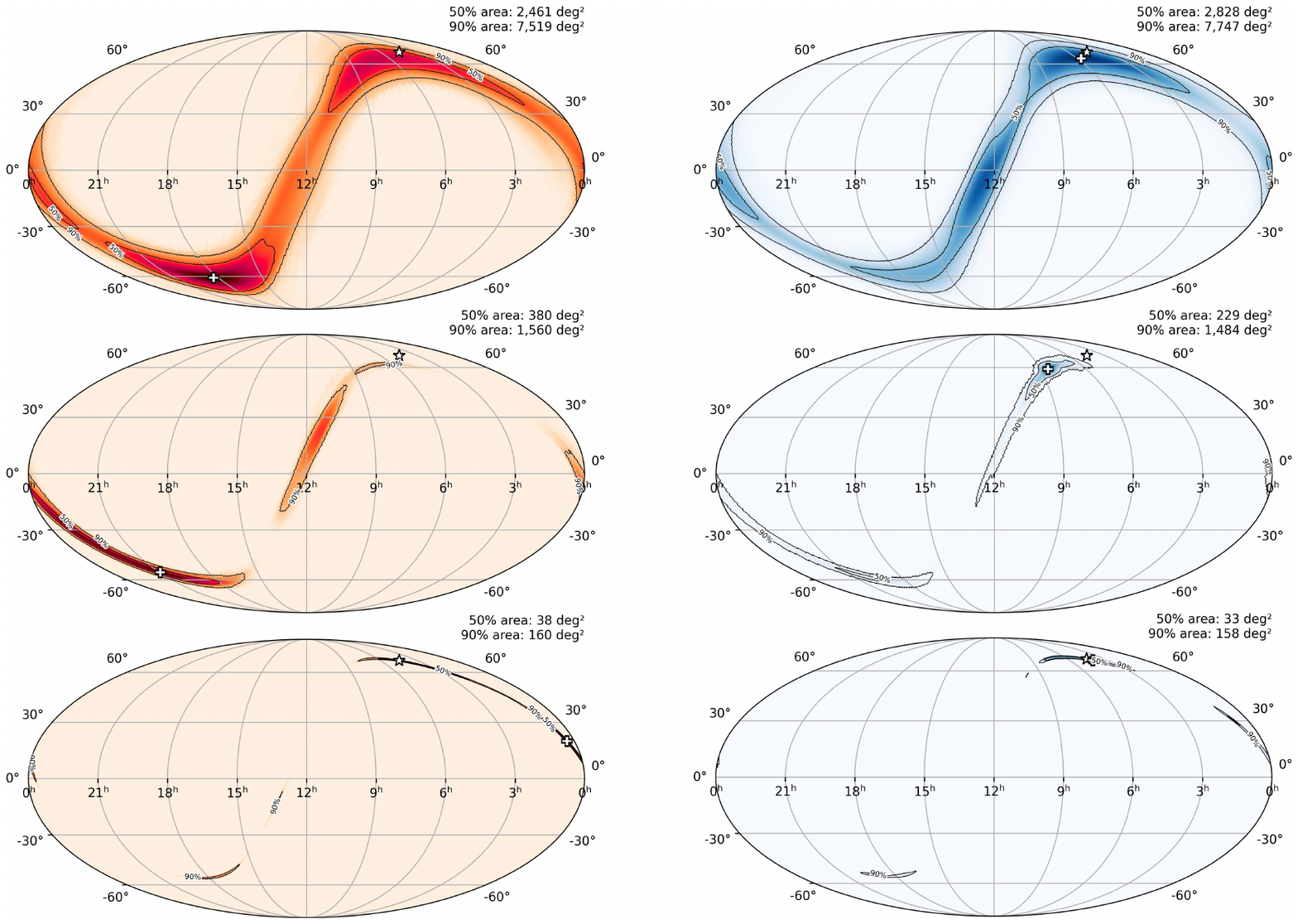}
\par\end{centering}
\caption{Illustration of early-warning skymaps for an eccentric injection. Here the injected eccentricity is 0.05 for a (2.76, 1.43) $M_{\odot}$ source represented with a star. The left and the right panels show the aligned-spin and the eccentricity-informed skymaps; top, middle and bottom panels show the skymaps obtained till 32, 64 and 512 Hz respectively. Given the initial probability densities of the aligned-spin skymaps (the most probable locations are shown with solid pluses), a telescope should begin tracking at the lower left part in the skymap. However, the telescope ideally needs to be repointed as the final skymap indicates to a different location in the sky. The eccentric localizations do not require significant repointings in this case and the most probable locations are relatively more accurate. Here the quoted confidence areas are largely similar in both types of skymaps but we note the existence of such an effect in addition to the cases where the localization areas are themselves numerically improved. A similar effect is reported for non-eccentric sources in~\citep{tohuvavohu2024swiftlychasinggravitationalwaves}.}\label{fig:ewskymaps}
\end{figure*}
\begin{figure*}[t]
\begin{centering}
\includegraphics[scale=0.65]{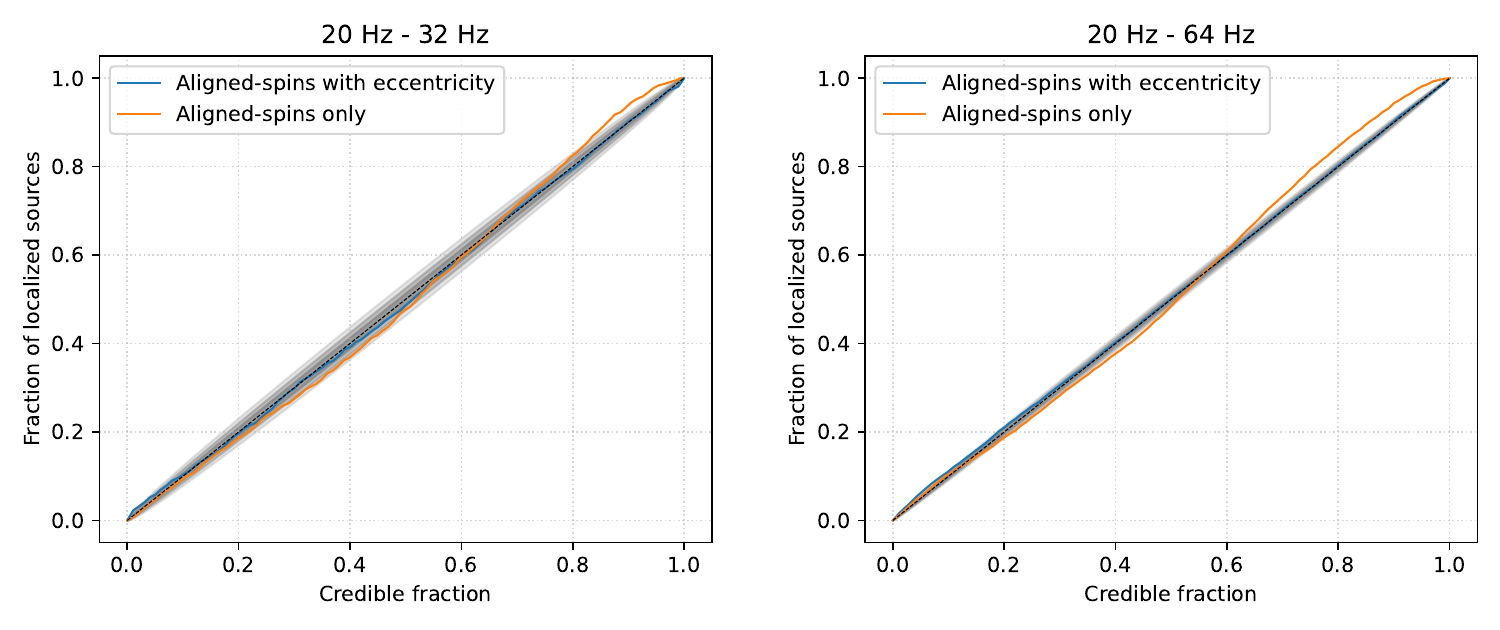}
\par\end{centering}
\caption{PP-plots obtained from the localization of commonly found eccentric injections by the EW searches. Plots for only the simulated O4 scenario are shown. Note that the current approach of localization is optimized toward the eccentric injections.}\label{fig:ppew}
\end{figure*}
\begin{center}
\begin{table}[b]
\begin{centering}
\begin{tabular}{>{\centering}p{2.5cm}>{\centering}p{1.8cm}>{\centering}p{1.8cm}>{\centering}p{1.8cm}}
\hline 
{\footnotesize{}Platform} & {\footnotesize{}32 Hz} & {\footnotesize{}64 Hz} & {\footnotesize{}512 Hz}\tabularnewline
\hline 
\hline 
{\footnotesize{}CPU} & {\footnotesize{}3.41 s} & {\footnotesize{}4.03 s} & {\footnotesize{}38.47 s}\tabularnewline
{\footnotesize{}GPU} & {\footnotesize{}1.21 s} & {\footnotesize{}2.24 s} & {\footnotesize{}...}\tabularnewline
\hline 
\end{tabular}
\par\end{centering}
\caption{Approximate runtimes for generating event candidates from 256 seconds of Gaussian HL data stretch using the spin-aligned eccentric PSO search, assuming that the data exist on the disk. The runtimes for CPU and GPU are obtained on a single \texttt{AMD EPYC 9554} and \texttt{NVIDIA Tesla V100} respectively utilizing all available cores. The GPU implementation is currently optimized for early-warning analyses only. The implementation of the early-warning localizations is not optimized and thus, have similar runtimes as for the full-bandwidth localizations.}\label{tab:hardware}
\end{table}
\par\end{center}
\begin{figure*}[t]
\begin{centering}
\includegraphics[scale=0.68]{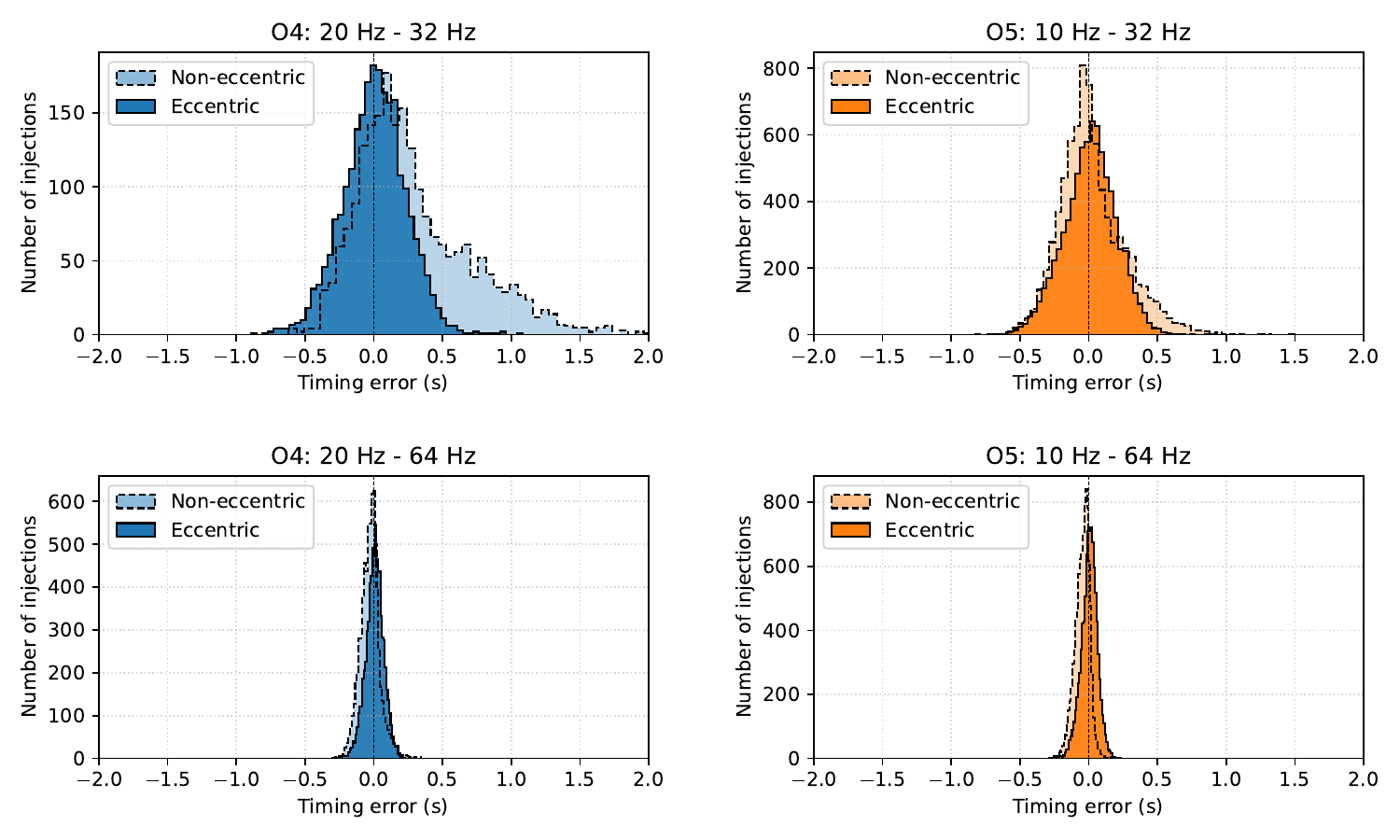}
\par\end{centering}
\caption{Difference between the injected $t_{c}$ of the simulated signals and the mean of the estimated arrival times at H1 and L1 obtained from the PSO based EW searches. The number of injections are out of the 10000 injections when the SNR of the triggers crossed 7 at both the detectors. It is highly unlikely that Gaussian noise can surpass this arbitrary threshold in any detector. The plot indicates that a coincident window of $\sim$ 15 ms is insufficient to capture all coincidences in the detector triggers of astrophysical origin.}\label{fig:ewtoa}
\end{figure*}
\section{Early-warning scenarios}\label{sec:earwar}
\begin{figure*}[t]
\begin{centering}
\includegraphics[scale=0.6]{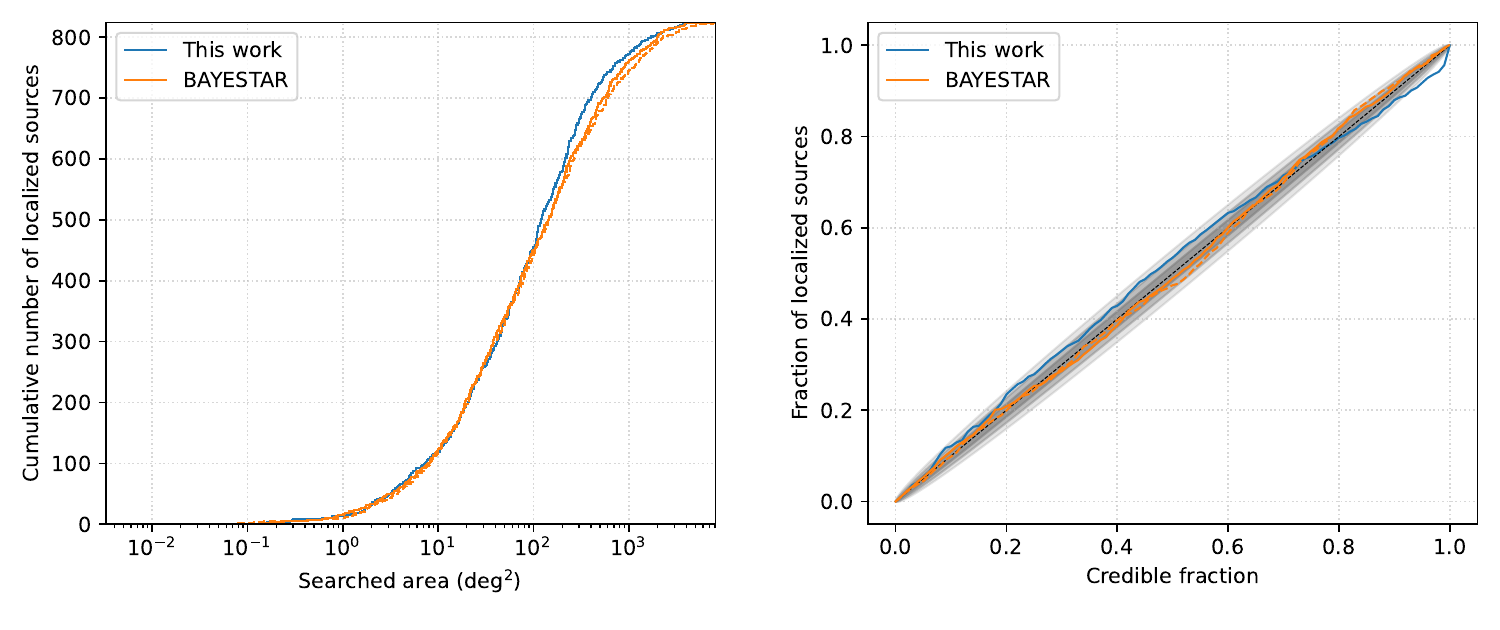}
\par\end{centering}
\caption{Accuracy (left) and self-consistency (right) of the localization areas compared for the simulated aligned-spin sources. We note that in the absence of eccentricity in the signals, the localizations accuracy of the current technique is comparable to that of \texttt{BAYESTAR}. Similar to Fig.~\ref{fig:prodsky}, the localization uncertainties are shown to be fairly self-consistent at 99.7\% confidence.}\label{fig:testsky}
\end{figure*}
In the previous section, we discussed the localization of simulated sources assuming that the entire in-band signal is present in the strain datasets. In this section, we describe an eccentric early-warning (EW) search by truncating these datasets so as to contain an incomplete inspiral without the merger. The basic idea behind an EW analysis is to provide a pre-indication of the merger of an incoming signal~\citep{Nitz_2020_1,Sachdev_2020,Kovalam_2022}. In general, an EW search has access to a smaller portion of signal and hence, the EW search processes a part of the full frequency bandwidth of the signal. Thus, an EW candidate is also expected to be followed by its regular full-bandwidth (FB) candidate counterpart. Usually, multiple frequency bands are chosen for EW searches that provide different EW times depending upon the high-frequency cutoff ($f\textsubscript{upper}$). For demonstration purposes, we choose only two such bands with $f\textsubscript{upper}$ of 32 Hz and 64 Hz. An $f\textsubscript{upper}$ of 32 Hz accumulates smaller SNRs but allows larger EW times while an $f\textsubscript{upper}$ of 64 accumulates larger SNRs but leaves smaller EW times. The low-frequency cutoff ($f\textsubscript{lower}$) is set for the low-frequency sensitivities of the instruments.

We generate a set of 10000 injections with the signal parameters given in Table.~\ref{tab:injpar} except that the luminosity distances here have an upper limit of 100 Mpc. These are deposited into Gaussian noise simulated for O4-like and O5-like sensitivities. They are recovered with the PSO-based searches with the similar procedure as described in Section~\ref{sec:simsea}, except that these are now conducted in the reduced frequency ranges. Here, to optimally recover the network SNRs, a smaller template-sampling is sufficient depending upon the frequency bandwidth. Thus, faster generation of templates truncated in the frequency space, the smaller data sampling rate and an optimized implementation of the PSO algorithm combined together enable considerable speedup. We have further implemented the analyses to be completed within a few seconds leveraging on GPUs. These are summarized in Table.~\ref{tab:hardware}.

The commonly found injections are then localized in a similar fashion as described in Section~\ref{sec:skyloc}. The accuracies of the localization areas are compared for the observing scenarios as shown in Fig.~\ref{fig:ear_loc}. We observe that the eccentric localization is consistently more accurate than the non-eccentric one in all cases. However, the improvements are generally more prominent in O5-like sensitivities. We primarily attribute this to the greater SNR gain from the numerous low-frequency inspiral cycles while considering eccentricity at the detection stage, as indicated in Fig.~\ref{fig:mchvsecc}. We argue that at very low data sampling rates, the SNR consistency across the detectors would carry relatively significant localization information, given the large errors in time and related phase measurements. An interesting comparison of the skymaps at different sampling frequencies is illustrated in Fig.~\ref{fig:ewskymaps}. The self-consistency of the sky areas obtained from the EW localizations are validated in Fig.~\ref{fig:ppew}.

In the above discussion, we have assumed a 15 ms time-window to test coincidences between triggers from LIGO Hanford (H1) and LIGO Livingston (L1). However, since the strain data for the EW analyses are critically downsampled, the timing uncertainty of the triggers of even astrophysical origin can result into non-coincidence, though they pass an SNR theshold criterion. This is demonstrated for the eccentric injections recovered with SNRs of at least 7 in both detectors as shown in Fig.~\ref{fig:ewtoa}. Note that the improved SNRs can lower the timing uncertainties in general, see for example~\citep{Fairhurst_2011}. We further observe that while an eccentric recovery obtains unbiased timing measurements on average, the aligned-spin only searches can report triggers with larger timing errors and/or with bias. Such errors are expected to propagate into the localizations which we do not explore here further. Refer to Appendix~\ref{sec:ewback} for investigations on the background event rates resulting from increased coincidence window.
\section{Conclusions}
In this work, we have described a novel, computationally optimized variant of the Particle Swarm Optimization (PSO) algorithm to search for compact binary mergers with eccentricity and large aligned-spins. It has the potential to deliver event candidates resulting from the search in near-realtime to facilitate immediate EM follow-up.

Through simulations, we have shown that the orbital eccentricity in inspiraling neutron stars can be used to improve their sky localization from gravitational wave observations. However, this necessitates that existing algorithms be optimized for detecting and localizing eccentric binaries. Here we have demonstrated a semi-Bayesian approach to conduct such localizations that uses the output of a matched-filter eccentric search. While we have discussed the implications for binary neutron stars, this approach can be easily extended to any modeled eccentric binary merger. The chances of observations also crtitically depend on the astrophysical abundancy of the sources. Even if few detectable sources exist within the reach of the detectors, current analyses can miss substantial SNRs and hence, the opportunity to pin down their sky positions. With improved low-frequency sensitivities, lighter binaries with greater early-inspiral cycles are relatively more prone to losing signal power. We have explored qualitative improvements with an eccentric early-warning system over a non-eccentric one with the current sensitivities of the detectors. We have identified some important cases with eccentric early-warning localizations from the EM-observations viewpoint.

While we have used only a two detector HL network for demonstration, it is useful to investigate the cases when more detectors are included such as Virgo and KAGRA. Given the semi-Bayesian nature of our approach, the localization method described here is expected to readily handle sub-threshold signals in any given detector(s). Additional detectors should guide the network's senstivity to the detectable binaries. Further improvements in localizing the astrophysical sources are awaited when LIGO India comes online~\citep{Saleem_2021,PhysRevD.109.044051}.

We have conducted studies on simulated data and provided only rough latency estimates without considering practical overheads. However, end-to-end tests of the capabilties of the eccentric early-warning system requires deployment of the search with a real-time interface. Fortunately, such an infrastructure is already operational within the LIGO-Virgo Collaboration~\citep{Magee_2021,Chaudhary_2024}. Since an eccentric EW search can accumulate greater SNRs in general, they can potentially reach a given significance threshold quicker, allowing greater EW times for eccentric sources. With the improved low-frequency senstivities of the instruments, we find that a dedicated eccentric EW system is desirable as discussed further in Appendix~\ref{sec:ewsea}. Such a system will improve the chances for the first and the subsequent EW observations. Though the localization algorithm presented in this work implements a computationally effective strategy, further optimizations are needed to practically realize EW observations. Alternately, current algorithms can be optimized to include eccentricity as an input parameter. Machine learning based approaches can already reach very low latencies~\citep{Chatterjeew_2023}.
\setlength\extrarowheight{0.5pt}
\begin{flushleft}
\begin{table*}[hbt!]
\vspace{0.35cm}
\begin{raggedright}
\centering
\begin{tabular}{>{\centering}m{3.0cm}>{\centering}m{3.0cm}>{\centering}m{3.0cm}>{\centering}m{1.1cm}>{\centering}m{3.0cm}>{\centering}m{3.0cm}}
\hline
\multirow{2}{2cm}{Configuration} & \multicolumn{2}{c}{Aligned-spin injections} &  & \multicolumn{2}{c}{Spin-aligned eccentric injections}\tabularnewline
\cline{2-3} \cline{3-3} \cline{5-6} \cline{6-6}
 & This work & \texttt{BAYESTAR} &  & This work & \texttt{BAYESTAR}\tabularnewline
\hline
\hline
Injected & ... & 0.9 &  & 0.5 & ...\tabularnewline
Estimated & 0.55 & 0.8 &  & 0.5 & 0.7\tabularnewline
\hline
\end{tabular}
\par\end{raggedright}
\begin{raggedright}
\caption{The log-likelihood ratio (or the SNR) rescaling factors used to obtain the pp-plots in Fig.~\ref{fig:prodsky} and Fig.~\ref{fig:testsky}.\label{tab:reslik}}
\par\end{raggedright}
\end{table*}
\par\end{flushleft}
\section{Acknowledgments}
We thank K. Rajesh Nayak for the providing valuable feedback on this work. We have used \texttt{ligo.skymap}, \texttt{PyCBC}, \texttt{dynesty}, \texttt{numpy}, \texttt{cupy}, and other packages for performing the analyses. Some of the results in this paper have been derived using the \texttt{healpy} and \texttt{HEALPix} package. The research and development work was carried out at the Center of Excellence in Space Sciences India (CESSI). CESSI is a multi-institutional Center of Excellence hosted by the Indian Institute of Science Education and Research (IISER) Kolkata and has been established through funding from the Ministry of Education, Government of India. We acknowledge the use of the Sarathi computing cluster at IUCAA for computational/numerical work. A portion of the GPU benchmarking was performed on the Kepler cluster hosted by the Department of Physical Sciences, IISER Kolkata. Partial support for the project was received from the Council for Scientific and Industrial Research (CSIR), India through File No:09/921(0272)/2019-EMR-I. This material is based upon work supported by NSF's LIGO Laboratory which is a major facility fully funded by the National Science Foundation. The article has a LIGO Document number LIGO-P2500001. We thank Gaurav Waratkar for carefully reading the manuscript. We sincerely acknowledge the valuable feedback from the anonymous referee.

This research has made use of data or software obtained from the Gravitational Wave Open Science Center (gwosc.org), a service of the LIGO Scientific Collaboration, the Virgo Collaboration, and KAGRA. This material is based upon work supported by NSF's LIGO Laboratory which is a major facility fully funded by the National Science Foundation, as well as the Science and Technology Facilities Council (STFC) of the United Kingdom, the Max-Planck-Society (MPS), and the State of Niedersachsen/Germany for support of the construction of Advanced LIGO and construction and operation of the GEO600 detector. Additional support for Advanced LIGO was provided by the Australian Research Council. Virgo is funded, through the European Gravitational Observatory (EGO), by the French Centre National de Recherche Scientifique (CNRS), the Italian Istituto Nazionale di Fisica Nucleare (INFN) and the Dutch Nikhef, with contributions by institutions from Belgium, Germany, Greece, Hungary, Ireland, Japan, Monaco, Poland, Portugal, Spain. KAGRA is supported by Ministry of Education, Culture, Sports, Science and Technology (MEXT), Japan Society for the Promotion of Science (JSPS) in Japan; National Research Foundation (NRF) and Ministry of Science and ICT (MSIT) in Korea; Academia Sinica (AS) and National Science and Technology Council (NSTC) in Taiwan.
\appendix
\section{Fallback localizations and the uncertainty estimates}\label{sec:locunc}
In the absence of eccentric sources, the algorithm retreats to the localization accuracies optimized for the aligned-spin case. To demonstrate, we generate 1000 injections from the same population as given in Table.~\ref{tab:injpar} but with $\varepsilon=0$. These are recovered with the aligned-spin PSO search. The recovered injections are localized using \texttt{BAYESTAR} and the localization algorithm introduced in this work. These are summaried in Fig.~\ref{fig:testsky}. The correction factors needed to obtain the pp-plots are obtained empirically, by systematically varying the value of the factor and checking the resulting pp-plots. The value at which a plot is reasonably consistent with the diagonal is noted for each case, as shown in Table.~\ref{tab:reslik}. Such factors are used to rescale the log-likelihood ratio while generating the posterior samples. We have generally observed that within a given method, the factor shifts more from unity when there is more deviation of the estimated parameters from the injected ones. This aspect has been discussed in detail in~\citep{PhysRevD.110.102002}. Note that the tunable parameter can be modified depending on the target population of sources and does not affect the accuracy of localizations. We do not expect these values to change considerably when the algorithm is used with real signals with the reasonable assumption that real instrument noise can be treated as approximately Gaussian except for the occasional instances of glitches localized in time.
\section{Distance estimates for eccentric injections}\label{sec:dist}
\begin{figure*}[t]
\begin{centering}
\includegraphics[scale=0.65]{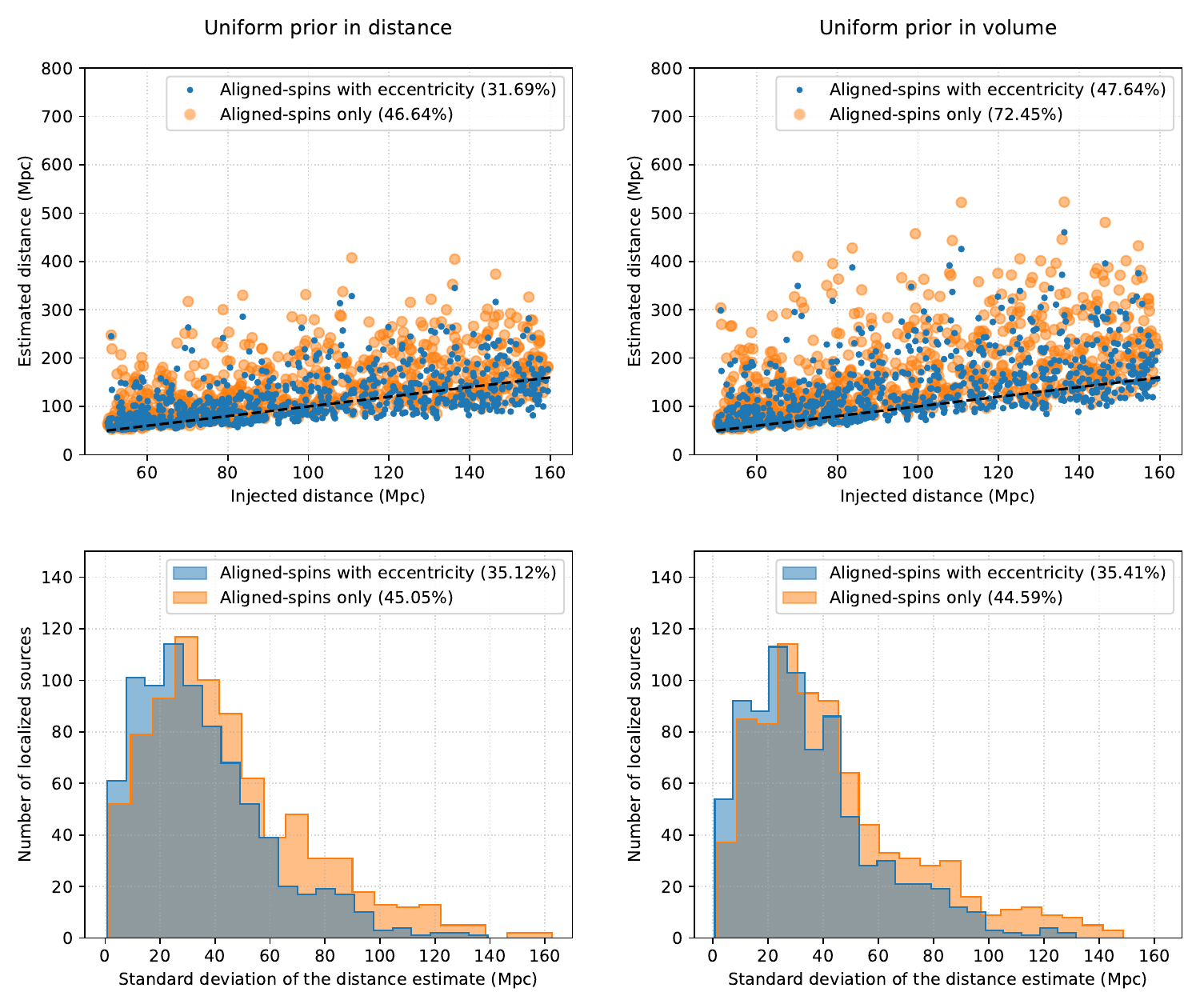}
\par\end{centering}
\caption{Summary of the distance estimates marginalized over the sky obtained under two different prior choices. We use a uniform prior for the luminosity distance (left) and a uniform prior in volume (right), determined by the sensitivity of the detectors, as described in the text. For these cases, a comparison of the mean errors (top) and the standard deviations (bottom) of the estimates is shown for the commonly found injections obtained with and without including eccentricity. The values in the parentheses indicate the average fractional error or the average standard deviation calculated over the localized sources.}\label{fig:distacc}
\end{figure*}
\begin{figure*}[t]
\begin{centering}
\includegraphics[scale=0.65]{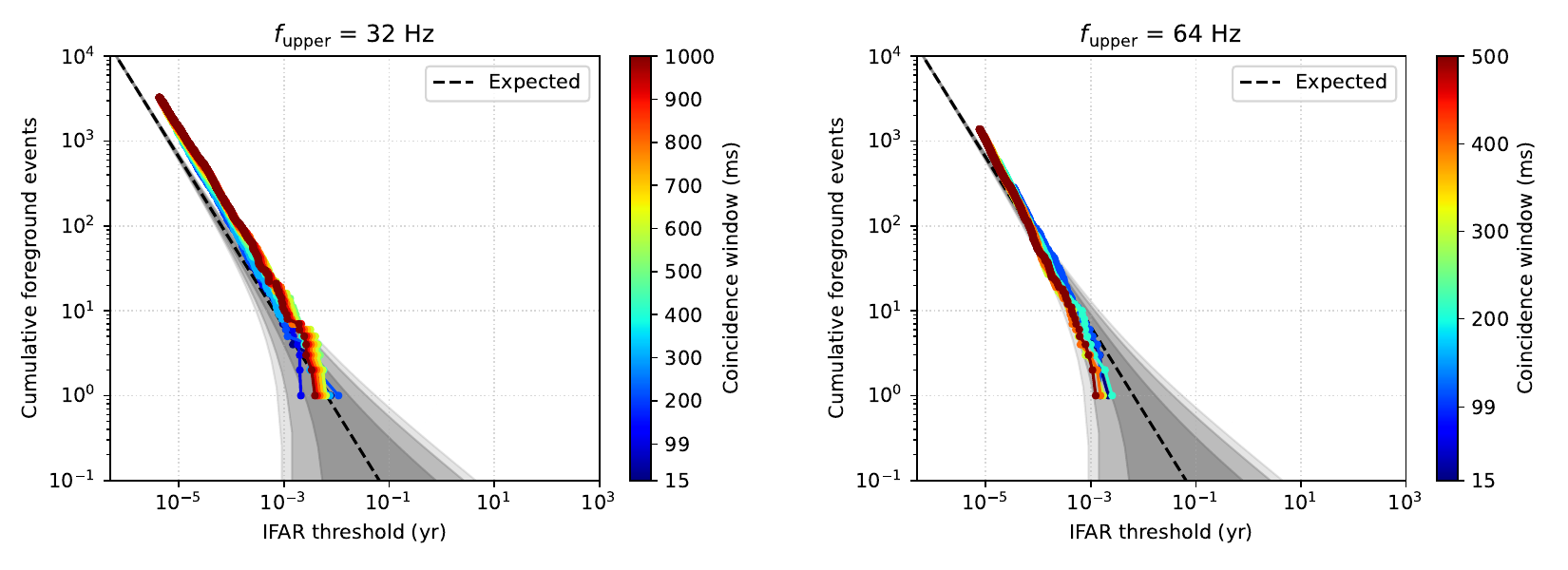}
\par\end{centering}
\caption{Self-consistency of the FAR estimates with varying coincidence windows investigated for the EW searches using datasets from O3b. The gray regions represent the $1$, $2$, and $3\sigma$ Poisson uncertainty estimates around the expected line. The largest coincidence window chosen till including fairly all injections shown in Fig.~\ref{fig:ewtoa}. Time-slide interval is twice the coincidence window.}\label{fig:ifarewlarge}
\end{figure*}
\begin{figure}[b]
\begin{centering}
\includegraphics[scale=0.6]{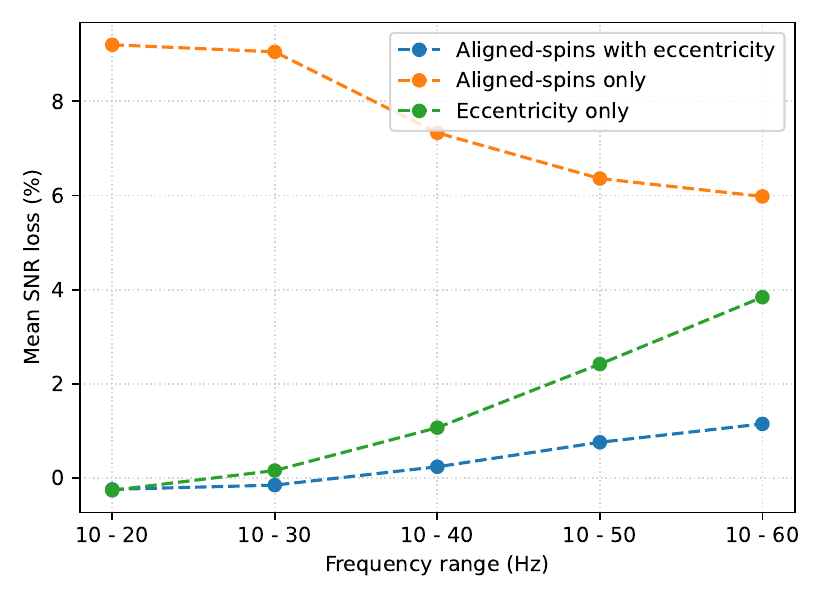}
\par\end{centering}
\caption{Average of the early-warning SNRs lost by various searches for the commonly recovered injections with O5-projected HL sensitivities as a function of the early-frequency bands.}\label{fig:ewfreq}
\end{figure}
\change{In a 3D localization, the reconstructed distance to a source is a function of the sky direction, i.e, there exists a unique posterior probability distribution for the distance along any given sky direction, in general. To assess the distance estimate, the distribution along the direction corresponding to the most probable pixel of a skymap can be used. However, the true sky position may not be be along a given inferred direction, so we use the distribution marginalized over all sky directions. The mean and the standard deviation of the distribution can be extracted from the header (\texttt{DISTMEAN} and \texttt{DISTSTD}) of the 3D \texttt{HEALPix} maps. These are described in~\citep{Singer_2016,Singer_2016_1}. Here we discuss the estimates obtained with the eccentric and the non-eccentric analyses under two different priors choices, as shown in Fig.~\ref{fig:distacc}. We observe that, on average, the errors in the estimates are smaller for the eccentric case than for the non-eccentric one. Further, the standard deviation of the estimates of the localized sources is generally smaller in the eccentric case. These are consistent across both the prior choices- a uniform prior in the distance itself and a uniform prior in the volume, each limited by a fiducial maximum reach of the detectors approximately determined by the simulated sensitivity. However, the errors are generally smaller for the uniform prior in distance that we attribute to the choice of the injection distribution, which is also uniform in distance. The use of astrophysically informed priors can be further explored in the future. We obtain negligible impact on the accuracies of the areal localizations due to the choice of these priors as long as the priors cover the injection space.}
\section{Backgrounds event rates in early-warning searches}\label{sec:ewback}
In this section, we explore the suitability of increasing the time-coincidence window for early-warning searches beyond the standard value used in the full-bandwidth searches. A coincidence window of 15 ms for the HL detector network can potentially miss several astrophysical EW candidates due to timing uncertainties of EW triggers, though, such events are expected to feature in an FB analysis. Thus, to improve the sensitivities of the EW analyses, a large coincidence window may be chosen. However, a larger coincidence window is also expected to allow more triggers of non-astrophysical origin to enter into coincidence. This would result into undesirable events, increasing false alarms. To investigate such chances for the PSO-based searches, we generate EW triggers in O3b data~\citep{RICHABBOTT2021100658,abbott2023open}. Here we use datasets from LIGO Hanford and LIGO Livingston to generate EW triggers from the same parameter space described earlier in this work. To compute the background events, the triggers from the individual detectors are time-slided with respect to each other. These processes are described in~\citep{pal2023swarm}. Here, the time-slide interval is assumed to be twice as large as a time-coincidence window used. We repeat this procedure with the same triggers but by varying width of time-window beyond 15 ms for the HL network. The maximum value used is motivated from the possible timing errors associated with a given frequency band estimated in Fig.~\ref{fig:ewtoa}. To accomodate triggers passing a smaller SNR threshold, a larger width for the coincidence window may be necessary. We observe that such triggers result into foreground coincidences with self-consistent inverse false alarm rate (IFAR) estimates, as shown in Fig.~\ref{fig:ifarewlarge}. Thus, we conclude that allowing for a larger coincidence window to accomodate a larger fraction of astrophysical signals does not interfere with their significance calculation in general.
\section{SNR loss in early-warning searches}\label{sec:ewsea}
Here we investigate the SNR loss from the PSO-based early-warning searches with several search-parameter combinations, in O5-like scenarios. The injection population is kept fixed as described earlier in Sec.~\ref{sec:earwar}. We use three combinations of the search parameters- (1) aligned-spins with eccentricity, (2) aligned-spins only, and (3) eccentricity only. The search configurations are otherwise kept same, including the template-sampling. For a similar discussion, refer to~\citep{Pal_2025}. The average SNR losses for the commonly found injections as a function of the EW frequency ranges are plotted in Fig.~\ref{fig:ewfreq}. We observe that the SNR loss by the EW search with both aligned-spin and eccentricity outperforms other searches at all frequency ranges. We also find that at very low frequency ranges (notably, in the 10-20 Hz), the non-spinning but eccentric search recovers SNRs closely to the search with both aligned-spins and eccentricity. Note the non-eccentric but aligned-spin search suffers maximum and significant loss in the SNR in this frequency range. This indicates that while eccentricity is important to be included, the aligned-spin components could possibly have only sub-dominant effects at such low frequencies.
\bibliography{ecc_sky}
\bibliographystyle{apsrev4-1}
\end{document}